\documentclass[prd]{revtex4}
\usepackage{epsfig}
\usepackage{graphicx}
\usepackage{dcolumn}
\usepackage{amsmath}
\usepackage{subfig}
\usepackage{latexsym}
\usepackage{hyperref}
\graphicspath{ {./Graphs/} }
\usepackage{wrapfig}

\begin{document}
\title{Effects of Gauss-Bonnet gravity on thermodynamics of (3+1)$-D$ AdS black holes}  

\author{Neeraj Kumar$^{a}$\footnote{e-mail:
neerajkumar@bose.res.in},Sunandan Gangopadhyay$^{a}$\footnote{e-mail: sunandan.gangopadhyay@bose.res.in,  ~~sunandan.gangopadhyay@gmail.com}}
\affiliation{$^a$Department of Theoretical Sciences,\\
S.N.Bose National Centre for Basic Sciences,\\
JD Block, Sector III, Salt Lake, Kolkata, 700106, India\\}
\maketitle
\section*{Abstract}
\noindent We studied the thermodynamics of the black holes with Gauss-Bonnet correction term in $3+1-$ dimensional AdS spacetime. It is known that the term has no effect on the equation of motion, however, it modifies the entropy formula of Wald as it has an effect of the Gauss-Bonnet parameter term. We studied charged black holes, namely, Reissner-Nordstr\"{o}m and Born-Infeld under this regime. We encountered divergences in heat capacity. After eliminating the possibility of first order phase transition, we applied two well trusted methods from standard thermodynamics, namely, Ehrenfest scheme and Ruppeiner's state space geometry analysis to ensure the second order nature of phase transition points. Effects of Gauss-Bonnet and Born-Infeld parameter are also discussed.
\section{Introduction}
\noindent Geometric definition of the black hole entropy is at the heart of black hole thermodynamics as it seems to resolve the discrepancy between thermodynamics and gravity. It was proposed by Bekenstein in \cite{bek1}, where entropy was associated to black holes using Hawking's area theorem \cite{hawk} as an analogy to preserve the second law of thermodynamics. Mere simplicity of black holes and their analogous behaviour to thermodynamic objects was shown in \cite{bhm} in the form of black hole mechanics. The analogy however was made robust by Hawking in \cite{hr} showing thermal nature of black holes using quantum effects of matter fields in curved spacetime. A lot of research has been going on since then to understand the thermodynamics of these objects with an aim to ultimately get to the quantum nature of gravity.\

\noindent There are many modifications to Einstein's general relativity theory even at classical level with a fair amount of motivation as general relativity has not yet been verified in strong limit. Lovelock theory \cite{lov} is one such higher curvature correction theory with higher curvature terms in the lagrangian, and when the action is modified up to second term it is named as Gauss-Bonnet gravity. Black hole solutions exist for these theories with the same symmetry as that of Schwarzchild black hole in Einstein's theory. Gauss-Bonnet term has no effect on the equation of motion for $D=4$ spacetime dimensions and the effects only appear for $D\geq5$. Wald has shown that in general diffeomorphism invariant theories, the entropy associated with a black hole turns out to be the Noether charge associated with the general coordinate transformations \cite{rm}. Wald's entropy for the black hole is given by the following expression
\begin{eqnarray}
S=2\pi\int_{\sum}d\phi\sqrt{g_{\phi\phi}}\dfrac{\partial L}{\partial R_{\mu\nu\lambda\rho}}\epsilon_{\mu\nu}\epsilon_{\lambda\rho}~.
\end{eqnarray}
 For Gauss-Bonnet black hole, entropy has a correction term apart from the area law depending on Gauss-Bonnet parameter as calculated in \cite{ent} and recently used in \cite{ent1}. It is interesting to see the effects of this correction to entropy on black hole thermodynamic properties. We wish to study charged black holes in Gauss-Bonnet gravity. We discuss two cases where black hole is charged due to Maxwell theory and Born-Infeld theory as it is always important to check the effects of non-linear electromagnetic theory on black holes. The thermodynamics of these black holes when cosmological constant treated as a thermodynamic variable is discussed in \cite{ent1} and gives rise to a pressure-volume term in the first law as first proposed in \cite{roy}. It is also interesting to check the behaviour when cosmological constant is not a thermodynamic variable and its value is fixed to some constant negative number. Results are known for black holes in Einstein gravity and it is known that the black holes undergo second order phase transition as discussed in \cite{r1}, \cite{r2}.\ So it is worth checking the effects of Gauss-Bonnet parameter in general on the nature of phase transition.

\noindent This paper is divided in two main parts. First part involves thermodynamic analysis of Reissner-Nordstr$\ddot{o}$m black holes. We calculated the thermodynamic quantities and find infinite divergences in heat capacity of the black hole when plotted with entropy. The Hawking temperature of the black hole is smooth function of entropy hence singular nature of heat capacity hints some phase transition of order higher than one. We analysed these singularities further using Ehrenfest scheme which has been used for black holes extensively in (\cite{r1}~-~\cite{n2})  and many more. Prigogine-Defay ratio as discussed in \cite{pr} is a parameter in standard thermodynamics giving deviation from second order phase transition is also calculated. Further, we also used Ruppeiner state space geometry analysis which again is a well trusted technique in standard thermodynamics as well as in black hole thermodynamics as discussed in \cite{ru1}, \cite{ru2}. Second part of the paper involves analysing same black hole in non-linear electromagnetic theory. We discussed Born-Infeld theory with an aim to understand the effects of non-linear parameter on thermodynamic properties as well as on phase transition. Along the way, we also calculated Smarr relation in both the cases keeping all dimensionful variables in the first law.\

\noindent This paper is organised in the following form. Section-II discusses Gauss-Bonnet-Reissner-Nordstr$\ddot{o}$m black hole involving calculations of thermodynamic quantities followed by Ehrenfest analysis and geometric analysis. Section-III contains the same analysis for Gauss-Bonnet-Born-Infeld black hole. We summarize our results in section-IV. The paper also contains an Appendix.  
\section{Gauss-Bonnet-Reissner-Nordstr$\ddot{O}$m black hole}
\subsection*{Thermodynamics and first law}
\noindent In this section, we define the black hole spacetime system we are interested in and calculate its thermodynamic properties. We consider charged black hole in $3+1-$ dimensional Gauss-Bonnet corrected gravity theory in AdS spacetime. We work in units where $G=h=c=1$.
We start from the following action 
\begin{eqnarray}
I=\dfrac{1}{16\pi}\int d^4x\sqrt{-g}[R-2\Lambda +\alpha L_{GB}+L(F)]
\label{a}
\end{eqnarray}
where Gauss-Bonnet lagrangian $L_{GB}=R^2-4R_{\gamma\delta}R^{\gamma\delta}+R_{\gamma\delta\lambda\sigma}R^{\gamma\delta\lambda\sigma}$ and charge is due to Maxwell term $L(F)=F^{\mu\nu}F_{\mu\nu}$. 
Here $\alpha$ is Gauss-Bonnet parameter and $\Lambda=-\dfrac{3}{l^2}$ ($l$ being AdS radius).

\noindent  The solution of the black hole spacetime in this theory reads 
\begin{eqnarray}
ds^2=-f(r)dt^2+\dfrac{1}{f(r)}dr^2+r^2d\Omega^2_{2}
\label{d}
\end{eqnarray}

\noindent with $f(r)$ being 
\begin{eqnarray}
f(r)=1-\dfrac{2M}{r}+\dfrac{Q^2}{r^2}+\dfrac{r^2}{l^2}
\label{e}
\end{eqnarray} where $M$ is the mass and $Q$ is the charge of the black hole. Note that the Gauss-Bonnet term does not show any effect on the equation of motion in $3+1-$ spacetime dimensions.\

\noindent Mass of the black hole can be written in terms of horizon radius ($r_+$) by using $f(r_+)=0$, so
\begin{eqnarray}
M=\dfrac{r_+}{2}+\dfrac{Q^2}{2r_+}+\dfrac{r_+^3}{2l^2}~.
\label{f}
\end{eqnarray}
Also the Hawking temperature of the black hole can be calculated as 
 
\begin{eqnarray}
T& = &\dfrac{1}{4\pi}\left(\dfrac{\partial f}{\partial r}\right)_{r_+} \nonumber\\
& = &\dfrac{1}{4\pi}\left(\dfrac{1}{r_+}-\dfrac{Q^2}{r_+^3}+\dfrac{3r_+}{l^2} \right)~.
\label{g}
\end{eqnarray}In the rest of our analysis, the AdS radius ($l$) will be kept unity unless stated otherwise.\

\noindent Now the first law of thermodynamics for black hole is
\begin{eqnarray}
dM=TdS+\Phi dQ
\label{k}~
\end{eqnarray}
which is analogous to the first law in standard thermodynamics
\begin{eqnarray}
dU=TdS-PdV
\end{eqnarray}
with the identification of the pressure $P$ to the negative of the electrostatic potential $\Phi$, the volume to the charge
$Q$ and the internal energy $U$ to the mass of the black hole $M$.\

\begin{figure}[!ht]
		\subfloat[\label{f1}]{{\includegraphics[scale=.7]{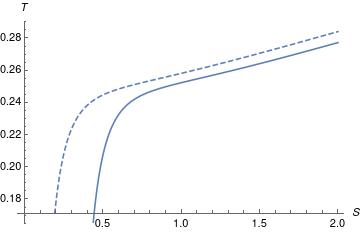}}}
		\qquad
		~~~~~~~~~~
		\caption{Temperature vs. Entropy}{Solid line (Q=0.2, $\alpha$=0.02)\\
		Dashed line (Q=0.2, $\alpha$=0)}
		\label{te}
\end{figure}

\noindent Therefore, the potential can be calculated as
\begin{eqnarray}
\Phi=\left(\dfrac{\partial M}{\partial Q}\right)_S=\dfrac{Q}{r_+}~.
\end{eqnarray}
\noindent Although the Gauss-Bonnet term does not affect the equation of motion, it changes the form of entropy and introduces an extra term apart from the area term. As calculated by using Wald formula in \cite{ent}, the entropy takes the form
\begin{eqnarray}
S=\pi(r_+^2+4\alpha)~.
\label{i}
\end{eqnarray}
So the Hawking temperature can be written in terms of entropy as
\begin{eqnarray}
T=\dfrac{1}{4\pi}\left(\sqrt{\dfrac{\pi}{S-4\alpha \pi}}-Q^2\left(\dfrac{\pi}{S-4\alpha \pi}\right)^{3/2}+3\sqrt{\dfrac{S-4\alpha \pi}{\pi}}\right)~.
\label{l}
\end{eqnarray}
Also the potential becomes
\begin{eqnarray}
\Phi=Q\sqrt{\dfrac{\pi}{S-4\pi\alpha}}~.
\label{ll}
\end{eqnarray}
\noindent We now plot the temperature with respect to entropy in Fig.(\ref{te}) for fixed values of charge and Gauss-Bonnet parameter. Temperature behaves smoothly with entropy and its behaviour in Einstein-Reissner-Nordstr$\ddot{o}$m limit (that is, $\alpha\rightarrow0$) is also plotted along to see the effects of the Gauss-Bonnet parameter.

 \noindent  Next, we calculate the heat capacity at constant potential $(C_{\Phi})$ and analyse the system thermodynamically.
Considering temperature of the black hole to be a function of entropy and charge  ($T\equiv T(S,Q)$), we can write
\begin{eqnarray}
\left(\dfrac{\partial T}{\partial S}\right)_{\Phi}=\left(\dfrac{\partial T}{\partial S}\right)_{Q}-\left(\dfrac{\partial T}{\partial Q}\right)_{S}\left(\dfrac{\partial \Phi}{\partial S}\right)_{Q}\left(\dfrac{\partial Q}{\partial \Phi}\right)_{S}~
\label{m}
\end{eqnarray}
where we use the thermodynamic identity
\begin{eqnarray}
\left(\dfrac{\partial Q}{\partial S}\right)_{\Phi}\left(\dfrac{\partial S}{\partial \Phi}\right)_{Q}\left(\dfrac{\partial \Phi}{\partial Q}\right)_{S}=-1
\label{n}
\end{eqnarray}
to write down eq.(\ref{n}).\

\noindent Using the thermodynamic expression in eq.(\ref{m}), the heat capacity at constant potential 
$C_{\Phi} =T\left(\dfrac{\partial S}{\partial T}\right)_{\Phi}$ can be calculated using eq.(s)(\ref{l},~\ref{ll})
\begin{eqnarray}
C_{\Phi}=2(S-4\pi \alpha)\left[\dfrac{\pi(S-4\pi\alpha)-Q^2\pi^2+3(S-4\pi \alpha)^2}{-\pi(S-4\pi\alpha)+Q^2\pi^2+3(S-4\pi \alpha)^2}\right]~~.
\label{o}
\end{eqnarray}
This reduces to the expression given in \cite{r1} when $\alpha=0$.
\begin{figure}[!ht]
		\subfloat[\label{f1}]{{\includegraphics[scale=.7]{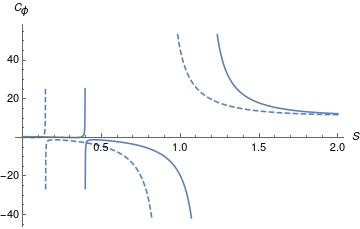}}}
		\qquad
		~~~~~~~~~~
		\caption{Heat Capacity vs. Entropy}{Solid line (Q=0.2, $\alpha$=0.02)\\
		Dashed line (Q=0.2, $\alpha$=0)}
		\label{hcc}
\end{figure}
\noindent We plot the heat capacity with entropy for same fixed parameters in Fig.(\ref{hcc}). It is apparant from the Figure that the black hole undergoes phase transition and exists in three different phases. One of the three phases also exhibit negative heat capacity. Since the Hawking temperature behaves smoothly with entropy so this eliminates the existence of first order phase transition and hints toward the possibility of second order phase transition. We shall be analysing this further. The effects of Gauss-Bonnet term on the phase transition points are also apparent from the plot. It shifts the phase transition points to higher entropy values. \

\noindent To end the discussion in this section, we calculate the Smarr relation for the black hole. In extended phase space thermodynamics, AdS radius $(l)$ has been shown to behave like thermodynamic pressure in \cite{roy} and is related to pressure as
\begin{eqnarray}
p=-\dfrac{\Lambda}{8\pi}=\dfrac{3}{8\pi l^2}~.
\label{ii}
\end{eqnarray}
Using eq.(s)(\ref{i}, \ref{ii}) in eq.(\ref{f}), we get the mass M to be
\begin{eqnarray}
M\equiv M(S,p,Q,\alpha)=\dfrac{1}{2}\sqrt{\dfrac{S-4\alpha \pi}{\pi}}+\dfrac{Q^2}{2}\sqrt{\dfrac{\pi}{S-4\alpha \pi}}+\dfrac{4\pi p}{3}\left(\dfrac{S-4\alpha \pi}{\pi}\right)^{3/2}~.
\label{iii}
\end{eqnarray}
Keeping all these dimensionful parameters in mind, the general form of first law of thermodynamics will take the form 
\begin{eqnarray}
dM=TdS+vdp+\Phi dQ+\Omega d\alpha
\label{j}
\end{eqnarray}
where $\Omega$ is some conjugate variable of Gauss-Bonnet parameter $\alpha$ and $v$ corresponds to $p$. These can be calculated from the expression of mass as $\Omega=\dfrac{\partial M}{\partial \alpha}$ and $v=\dfrac{\partial M}{\partial p}$. The dimensions of the quantities involved in eq.(\ref{iii}) are $[M]=[L],~[S]=[L]^2,~ [p]=[L]^{-2}, ~[Q]=[L]$ and $[\alpha]=[L]^2$. So the Smarr relation for the black hole can be calculated and it becomes
\begin{eqnarray}
M=2TS-2pv+Q\Phi +2\Omega \alpha~.
\end{eqnarray}
\noindent Here we can explicitly see that the Gauss-Bonnet parameter enters into the Smarr formula in $3+1-$ spacetime dimensions.

\noindent We shall now look at the phase transition points of the black hole indicated in the heat capacity plots. We shall analyse the order of phase transition using Ehrenfest scheme and Ruppeiner's state space geometry analysis in the following sections.

\subsection*{Analysis of phase transition using Ehrenfest scheme}
\noindent Ehrenfest's formalism of studying phase transitions is the standard technique in thermodynamics to determine the nature of phase transitions for various thermodynamical systems. It simply says that the order of the phase transition corresponds to the discontinuity in the order of the derivative of Gibb's potential. For second order phase transition the second derivative encounters discontinuities. However, first derivative and Gibb's potential at those points are continuous. These conditions with Maxwell's relations give two equations which have to be satisfied for second order phase transition.\

\noindent The first and second Ehrenfest equations in thermodynamics are given by
\begin{eqnarray}
\left(\dfrac{\partial P}{\partial T}\right)_S&=&\dfrac{1}{VT}\dfrac{C_{P_2}-C_{P_1}}{\beta_2-\beta_1}=\dfrac{\Delta C_P}{VT\Delta\beta},
\label{u}
\end{eqnarray}
\begin{eqnarray}
\left(\dfrac{\partial P}{\partial T}\right)_V&=&\dfrac{\beta_2-\beta_1}{\kappa_2-\kappa_1}=\dfrac{\Delta\beta}{\Delta\kappa}
\label{v}
\end{eqnarray}
where subscripts $1$ and $2$ denote two distinct phases of the system. Now we use the correspondence between the pressure ($P$) to the negative of the electrostatic potential difference ($\Phi$) and the volume ($V$) to the charge ($Q$) of the black hole. These identifications lead to the following equations 
\begin{eqnarray}
-\left(\dfrac{\partial \Phi}{\partial T}\right)_S&=&\dfrac{1}{QT}\dfrac{C_{{\Phi}_2}-C_{\Phi_1}}{\beta_2-\beta_1}=\dfrac{\Delta C_{\Phi}}{QT\Delta\beta}
\label{w}
\end{eqnarray}
\begin{eqnarray}
-\left(\dfrac{\partial \Phi}{\partial T}\right)_Q&=&\dfrac{\beta_2-\beta_1}{\kappa_2-\kappa_1}=\dfrac{\Delta\beta}{\Delta\kappa}~.
\label{x}
\end{eqnarray}
Note that $\beta$ is the volume expansion coefficient and $\kappa$ is the isothermal compressibility of the system and are defined as
\begin{eqnarray}
\beta=\dfrac{1}{Q}\left(\dfrac{\partial Q}{\partial T}\right)_{\Phi}~,~~~~
\kappa=\dfrac{1}{Q}\left(\dfrac{\partial Q}{\partial \Phi}\right)_{T}~.
\label{y}
\end{eqnarray}

\noindent Now we proceed to check whether the black hole phase transition satisfies the Ehrenfest equations (\ref{w}, \ref{x}). In other words, we investigate the validity of the Ehrenfest equations at the points of discontinuities $S_i$,~(i=1, 2). Here we denote the critical values of temperature by $T_i$ and charge by $Q_i$.
The left hand side of the first equation for the black hole spacetime takes the form
\begin{eqnarray}
-\left[\left(\dfrac{\partial \Phi}{\partial T}\right)_{S}\right]_{S=S_{i}}=\dfrac{2\pi}{Q_i}\left(\dfrac{S_i-4\alpha \pi}{\pi}\right)~.
\label{a1}
\end{eqnarray}
In getting this result, we have used 
\begin{eqnarray}
\left[\left(\dfrac{\partial \Phi}{\partial Q}\right)_{S}\right]_{S=S_i}=\sqrt{\dfrac{\pi}{S_i-4\pi\alpha}}~,
\label{new1}
\end{eqnarray}
\begin{eqnarray}
\left[\left(\dfrac{\partial T}{\partial Q}\right)_{S}\right]_{S=S_{i}}=-\dfrac{Q_i}{2\pi}\left(\dfrac{\pi}{S_i-4\pi\alpha}\right)^{3/2}
\end{eqnarray}
and eq.(\ref{z}) in the Appendix.
Also, the right hand side becomes
\begin{eqnarray}
\dfrac{\Delta C_{\Phi}}{T_i Q_i\Delta\beta}=\dfrac{2\pi}{Q_i}\left(\dfrac{S_i-4\alpha \pi}{\pi}\right)~
\label{a5}
\end{eqnarray}
which is computed using eq.(\ref{new1}), 
\begin{eqnarray}
\left[\left(\dfrac{\partial \Phi}{\partial S}\right)_{Q}\right]_{S=S_{i}}=-\dfrac{\sqrt{\pi} Q}{2}\left(\dfrac{1}{S_i-4\pi\alpha}\right)^{3/2}
\label{new2}
\end{eqnarray}
and eq.(\ref{a4}) in the Appendix.
\noindent Eq.(s)(\ref{a1},~\ref{a5}) show the validity of first Ehrenfest's equation for the black hole spacetime under consideration. We now proceed to check the second equation.\

\noindent Using eq.(\ref{a7}), the left hand side of second Ehrenfest equation becomes 

\begin{eqnarray}
-\left[\left(\dfrac{\partial \Phi}{\partial T}\right)_Q\right]_{S=S_{i}}=\dfrac{2\pi}{Q_i}\left(\dfrac{S_i-4\alpha \pi}{\pi}\right).
\label{a9}
\end{eqnarray}

\noindent The right hand side of the equation becomes
\begin{eqnarray}
\dfrac{\Delta \beta}{\Delta \kappa}=-\dfrac{4\pi^2Q_i(S_i-4\alpha \pi)}{-\pi(S_i-4\alpha\pi)+3Q_i^2\pi^2+3(S_i-4\alpha\pi)^2}~.
\label{a113}
\end{eqnarray}
Here we used eq.(\ref{new2}), the following expression
\begin{eqnarray}
\left[\left(\dfrac{\partial T}{\partial S}\right)_Q\right]_{S=S_i}=\dfrac{1}{4\pi}\left(-\sqrt{\pi}\left(\dfrac{1}{S_i-4\pi\alpha}\right)^{3/2}+\dfrac{3\pi^{3/2}Q_i^2}{2}\left(\dfrac{1}{S_i-4\pi \alpha}\right)^{5/2}+\dfrac{3}{2\sqrt{\pi}}\sqrt{\left(\dfrac{1}{S_i-4\pi\alpha}\right)}\right)
\end{eqnarray}
and eq.(\ref{a13}) in the Appendix.
Since the heat capacity has infinite discontinuities, therefore, eq.(\ref{o}) at points of phase transition gives the relation
\begin{eqnarray}
-\pi(S_i-4\alpha\pi)+Q_i^2\pi^2+3(S_i-4 \alpha\pi)^2=0
\label{a1133}
\end{eqnarray} 
Using above condition in eq.(\ref{a113}), we obtain the right hand side of second Ehrenfest equation to be
\begin{eqnarray}
\dfrac{\Delta \beta}{\Delta \kappa}=-\dfrac{2\pi}{Q_i}\left(\dfrac{S_i-4\alpha\pi}{\pi}\right)~.
\label{a16}
\end{eqnarray}
Eq.(s)(\ref{a9},~\ref{a16}) show the validity of second Ehrenfest equation at the critical points.\

\noindent We calculate the Prigogine-Defay (PD) ratio \cite{pr} for the system. Using eq.(s)(\ref{w},~\ref{a7}), we have
\begin{eqnarray}
\left[\left(\dfrac{\partial \Phi}{\partial T}\right)_Q\right]_{S=S_i}=\left[\left(\dfrac{\partial \Phi}{\partial T}\right)_S\right]_{S=S_i}=-\dfrac{\Delta C_{\Phi}}{T_i Q_i\Delta\beta}
\label{a17}
\end{eqnarray}
and eq.(\ref{a12}) with above equation gives
\begin{eqnarray}
\Pi=\dfrac{\Delta C_{\Phi}\Delta\kappa}{T_i Q_i(\Delta\beta)^2}=1~.
\label{a18}
\end{eqnarray}
These results show the exact second order nature of phase transition. Further we apply Ruppeiner's state space geometry analysis to reassure second order nature of phase transition.
\section*{State space geometry analysis}
\noindent Ruppeiner's state space geometry technique is another trusted mathematical formalism to study thermodynamic systems. The idea is to write the abstract manifold with the notion of distance from thermodynamic variables and study the curvature in order to study phase transition. Singularities in the manifold correspond to phase transition points. Definitions of metric coefficients are quite standard \cite{ru1}, \cite{ru2}.\

\begin{figure}[!ht]
		\subfloat[\label{f1}]{{\includegraphics[scale=.7]{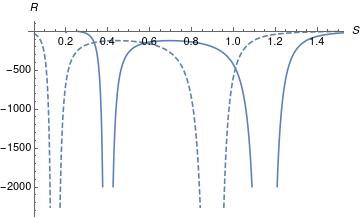}}}
		\qquad
		~~~~~~~~~~
		\caption{Ruppeiner curvature vs. Entropy}{Solid line (Q=0.2, $\alpha$=0.02)\\
		Dashed line (Q=0.2, $\alpha$=0)}
		\label{fiR}
\end{figure}

\noindent The Ruppeiner metric coefficients for the manifold are given by
\begin{eqnarray}
g^{R}_{ij}=-\dfrac{\partial^2 S(x^i)}{\partial x^i \partial x^j}
\label{a19}
\end{eqnarray}
where $x^i=x^i(M, Q)$; i=1,2 are extensive variables of the manifold.
\noindent The calculation of the Weinhold metric coefficients is convenient for computational purpose. These are given by
\begin{eqnarray}
g^{W}_{ij}=\dfrac{\partial^2 M(x^i)}{\partial x^i \partial x^j}
\label{a20}
\end{eqnarray}
where $x^i=x^i(S, Q)$; i=1,2. It is to be noted that Weinhold geometry is connected to the Ruppeiner geometry through the following map
\begin{eqnarray}
dS^2_R=\dfrac{dS^2_W}{T}~.
\label{a21}
\end{eqnarray}
By putting AdS radius to unity (that is, $p=\dfrac{3}{8\pi}$ from (\ref{ii})) in eq.(\ref{iii}), we get
\begin{eqnarray}
M=\dfrac{1}{2}\sqrt{\dfrac{S-4\alpha \pi}{\pi}}+\dfrac{Q^2}{2}\sqrt{\dfrac{\pi}{S-4\alpha \pi}}+\dfrac{1}{2}\left(\dfrac{S-4\alpha \pi}{\pi}\right)^{3/2}~.
\end{eqnarray}
Ruppeiner metric coefficients are calculated from the above equation and these take the form
\begin{eqnarray}
g_{QQ}&=&\dfrac{1}{T}\sqrt{\dfrac{\pi}{S-4\alpha\pi}}\nonumber\\
g_{SQ}&=&-\dfrac{1}{T}\dfrac{Q}{2(S-4\alpha\pi)}\sqrt{\dfrac{\pi}{S-4\alpha\pi}}\nonumber\\
g_{SS}&=&\dfrac{1}{8T}\sqrt{\dfrac{\pi}{S-4\alpha\pi}}\left[-\dfrac{1}{\pi(S-4\alpha\pi)}+\dfrac{3Q^2}{(S-4\alpha\pi)^2}+\dfrac{3}{\pi^2}\right]~.
\label{a22}
\end{eqnarray}
With these metric coefficients, we now calculate the standard curvature of the two dimensional manifold. Singularities in the curvature indicate second order phase transitions. The expression for the curvature in terms of the metric coefficients read \cite{fobi} \
\begin{eqnarray}
R =-\dfrac{1}{\sqrt{g}}\left[\dfrac{\partial}{\partial S}\left(\dfrac{g_{SQ}}{\sqrt{g}g_{SS}}\dfrac{\partial g_{SS}}{\partial Q}-\dfrac{1}{\sqrt{g}}\dfrac{g_{QQ}}{\partial S}\right)
+\dfrac{\partial}{\partial Q}\left(\dfrac{2}{\sqrt{g}}\dfrac{\partial  g_{SQ}}{\partial Q}-\dfrac{1}{\sqrt{g}}\dfrac{\partial g_{SS}}{\partial Q}-\dfrac{g_{SQ}}{\sqrt{g}g_{SS}}\dfrac{\partial g_{SS}}{\partial S}\right)\right]~.
\label{a27}
\end{eqnarray}
where $g=g_{SS}~ g_{QQ}-g_{SQ}^2$.
The analytical expresion is too large to write. We plot it with entropy for fixed value of other parameters in Fig.(\ref{fiR}).
From the graph we find that the Ruppeiner curvature diverges for all values where heat capacity is singular. This confirms the second order nature of phase transition. Same behaviour of Gauss-Bonnet parameter is observed through Ruppeiner curvature as well, as the points of phase transition are observed to shift to the right (that is, towards higher entropy value). In the following sections, we shall be doing the same analysis for Gauss-Bonnet black hole with Born-Infeld term in order to observe the non-linear effects of electromagnetism on the thermodynamic quantities.
\section{Gauss-Bonnet-Born-Infeld Black hole}
\subsection*{Thermodynamics}
\noindent In this section, we shall replace the Maxwell term from eq.(\ref{a}) with Born-Infeld term $L(F)=4b^2\left( 1-\sqrt{1+\dfrac{F^{\mu\nu}F_{\mu\nu}}{2b^2}}\right)$, where $b$ is Born-Infeld paramter. This lagrangian term reduces to Maxwell electrodynamics in the limit $b\rightarrow \infty $. Again, the Gauss-Bonnet term is not going to affect the equation of motion. In this case, the function $f(r)$ takes the form
\begin{eqnarray}
f(r)=1-\dfrac{2M}{r}+\dfrac{r^2}{l^2}+\dfrac{2b^2r^2}{3}\left(1-\sqrt{1+\dfrac{Q^2}{b^2r^4}}\right)+\dfrac{4Q^2}{3r^2}{}_2F_1\left[\dfrac{1}{4},\dfrac{1}{2},\dfrac{5}{4},-\dfrac{Q^2}{b^2r^{4}}\right]
\label{bo1}
\end{eqnarray}
where ${}_2F_1\left[\mu,\nu,\rho,x\right]$ is hypergeometric function.\

\noindent Mass of the black hole in terms of horizon radius ($r_+$) (calculated using $f(r_+)=0$) 
is 
\begin{eqnarray}
M=\dfrac{r_+}{2}+\dfrac{r_+^3}{2l^2}+\dfrac{b^2r_+^3}{3}\left(1-\sqrt{1+\dfrac{Q^2}{b^2r_+^4}}\right)+\dfrac{2Q^2}{3r_+}{}_2F_1\left[\dfrac{1}{4},\dfrac{1}{2},\dfrac{5}{4},-\dfrac{Q^2}{b^2r_+^{4}}\right]
\label{bo2}
\end{eqnarray}
 and the Hawking temperature is
 \begin{eqnarray}
 T=\dfrac{1}{4\pi}\left[\dfrac{1}{r_+}+\dfrac{3r_+}{l^2}+2b^2r_+\left(1-\sqrt{1+\dfrac{Q^2}{b^2r_+^4}}\right)\right]~.
 \label{bo3}
 \end{eqnarray}
\noindent Using first law of black hole thermodynamics from eq.(\ref{k}), we calculate the electrostatic potential at horizon to be
 \begin{eqnarray}
 \Phi=\left(\dfrac{\partial M}{\partial Q}\right)=\dfrac{Q}{r_+}{}_2F_1\left[\dfrac{1}{4},\dfrac{1}{2},\dfrac{5}{4},-\dfrac{Q^2}{b^2r_+^{4}}\right]~.
\label{bo4}
 \end{eqnarray}
 Since, Wald's entropy formula does not depend on the matter lagrangian, so it is independent of the Born-Infeld term and its expression is same as the one given in eq.(\ref{i}). Again, we shall put AdS radius $l$ to unity.\
 
\begin{figure}[!ht]
		\subfloat[\label{f1}]{{\includegraphics[scale=.7]{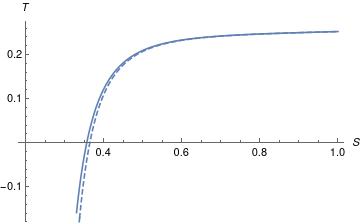}}}
		\qquad
		~~~~~~~~~~
		\caption{Temperature vs. Entropy}{Solid line(Q=0.2, $\alpha$=0.02, b=10)\\
		Dashed line (Q=0.2, $\alpha$=0.02, b$\rightarrow\infty$)}
		\label{vvv}
\end{figure}

\noindent The Hawking temperature and potential in terms of entropy read 
\begin{eqnarray}
T=\dfrac{1}{4\pi}\left[\sqrt{\dfrac{\pi}{S-4\alpha \pi}}+3\sqrt{\dfrac{S-4\alpha\pi}{\pi}}+2b^2\sqrt{\dfrac{S-4\alpha\pi}{\pi}}\left(1-\sqrt{1+\dfrac{Q^2\pi^2}{b^2(S-4\alpha\pi)^2}}\right)\right]
\label{BT}
\end{eqnarray}

\begin{eqnarray}
\Phi=Q\sqrt{\dfrac{\pi}{S-4\alpha\pi}}{}_2F_1\left[\dfrac{1}{4},\dfrac{1}{2},\dfrac{5}{4},-\dfrac{Q^2\pi^2}{b^2(S-4\alpha\pi)^2}\right]~.
\label{BP}
\end{eqnarray}
\noindent We plot the Hawking temperature with entropy in Fig.(\ref{vvv}) in order to see its thermodynamic behaviour. It is a smooth function of entropy.
The effect of Born-Infeld parameter can also be seen from the plot for same value of Gauss-Bonnet parameter. Black hole with non-linear electromagnetic term in Gauss-Bonnet gravity will have higher temperature than with Maxwell term for same entropy value and this effect is eminent at low entropy values.  \

\begin{figure}[!ht]
		\subfloat[\label{f1}]{{\includegraphics[scale=.7]{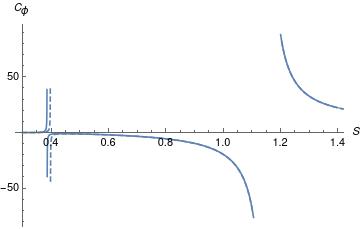}}}
		\qquad
		~~~~~~~~~~
		\caption{Heat Capacity vs. Entropy}{Solid line(Q=0.2, $\alpha$=0.02, b=10)\\
		Dashed line(Q=0.2, $\alpha$=0.02, b$\rightarrow\infty$) }
		\label{cck}
\end{figure}

\noindent Next, we calculate the heat capacity from eq.(\ref{m}):
\begin{eqnarray}
C_{\Phi} &=&T\left(\dfrac{\partial S}{\partial T}\right)_{\Phi}\nonumber\\
&=&\dfrac{T\left(\dfrac{\partial \Phi}{\partial Q}\right)_{S}}{\left(\dfrac{\partial \Phi}{\partial Q}\right)_{S}\left(\dfrac{\partial T}{\partial S}\right)_Q-\left(\dfrac{\partial T}{\partial Q}\right)_{S}\left(\dfrac{\partial \Phi}{\partial S}\right)_Q}~.
\label{ooo}
\end{eqnarray}
These derivatives can be calculated using eq.(s)(\ref{BT},~\ref{BP}) as
\begin{eqnarray}
\left(\dfrac{\partial \Phi}{\partial S}\right)_{Q}=-\dfrac{Q}{2(S-4\alpha \pi)}\sqrt{\dfrac{\pi}{S-4\alpha\pi}}\left(1+\dfrac{Q^2\pi^2}{b^2(S-4\alpha\pi)^2}\right)^{-1/2}~,
\label{df1}
\end{eqnarray}
\begin{eqnarray}
\left(\dfrac{\partial \Phi}{\partial Q}\right)_{S}=\dfrac{1}{2}\sqrt{\dfrac{\pi}{S-4\alpha\pi}}\left[\left(1+\dfrac{Q^2\pi^2}{b^2(S-4\alpha\pi)^2}\right)^{-1/2}+{}_2F_1\left[\dfrac{1}{4},\dfrac{1}{2},\dfrac{5}{4},-\dfrac{Q^2\pi^2}{b^2(S-4\alpha\pi)^2}\right]\right]~,
\label{df2}
\end{eqnarray}
\begin{eqnarray}
\left(\dfrac{\partial T}{\partial Q}\right)_{S}=-\dfrac{Q}{2(S-4\alpha \pi)}\sqrt{\dfrac{\pi}{S-4\alpha\pi}}\left(1+\dfrac{Q^2\pi^2}{b^2(S-4\alpha\pi)^2}\right)^{-1/2}~
\label{df3}
\end{eqnarray}and
\begin{eqnarray}
\left(\dfrac{\partial T}{\partial S}\right)_Q=\dfrac{1}{4\pi}\sqrt{\dfrac{\pi}{S-4\alpha\pi}}\left[-\dfrac{1}{2(S-4\alpha\pi)}+\dfrac{3}{2\pi}+\dfrac{b^2}{\pi}\left(1-\sqrt{1+\dfrac{Q^2\pi^2}{b^2(S-4\alpha\pi)^2}}\right)\right.\nonumber\\
\left.+\dfrac{2Q^2\pi}{(S-4\alpha\pi)^2}\left(1+\dfrac{Q^2\pi^2}{b^2(S-4\alpha\pi)^2}\right)^{-1/2}\right]~.
\label{df4}
\end{eqnarray}
We plot the heat capacity in Fig.(\ref{cck}).
Black hole undergoes phase transition at two points of entropy and exists in three different phases. Smooth T-S graph in Fig.(\ref{vvv}) excludes any first order phase transition, so there is again the possibility of second order phase transition. The qualitative behaviour of heat capacity is same as discussed for Einstein-Born-Infeld black holes in \cite{r2}. Non-linear electromagnetic term shifts the phase transition points to left. This effect is prominent at lower entropy values and it dies down as entropy increases as observed for the Hawking temperature as well.  \

\noindent We now calculate the Smarr relation of the black hole.   
 Using eq.(s)(\ref{i},\ref{ii}) in eq.(\ref{bo2}), we can write the mass as
 \begin{eqnarray}
 M\equiv M(S,p,Q,b,\alpha)=\dfrac{1}{2}\sqrt{\dfrac{S-4\alpha\pi}{\pi}}+\dfrac{1}{2l^2}\left(\dfrac{S-4\alpha\pi}{\pi}\right)^{3/2}+\dfrac{b^2}{3}\left(\dfrac{S-4\alpha\pi}{\pi}\right)^{3/2}\left(1-\sqrt{1+\dfrac{Q^2\pi^2}{b^2(S-4\alpha\pi)^2}}\right)\nonumber\\
 +\dfrac{2Q^2}{3}\sqrt{\dfrac{\pi}{S-4\alpha\pi}}{}_2F_1\left[\dfrac{1}{4},\dfrac{1}{2},\dfrac{5}{4},-\dfrac{Q^2\pi^2}{b^2(S-4\alpha\pi)^2}\right]~.
 \label{bo5}
 \end{eqnarray}
General first law of thermodynamics for the system becomes
\begin{eqnarray}
dM=TdS+vdp+\Phi dQ+\Omega d\alpha+Bdb
\end{eqnarray}
where $B$ is conjugate variable of $b$ and can be calculated as $B=\dfrac{\partial M}{\partial b}$. By dimensional analysis, we see the dimensions of the variables involved in the mass expression in eq.(\ref{bo4}) are $[M]=[L],~[S]=[L]^2,~ [p]=[L]^{-2}, ~[Q]=[L], ~[b]=[L]^{-1}$ and $[\alpha]=[L]^2$. Smarr relation for the black hole becomes
\begin{eqnarray}
M=2TS-2pv+Q\Phi -bB +2\Omega \alpha~.
\end{eqnarray}
\noindent Again we can see that the Gauss-Bonnet and the Born-Infeld parameter enter into the Smarr formula in $3+1-$ spacetime dimensions. 

\noindent Next, we turn to phase transition points and their order. We check the status of Ehrenfest equations for the system and also apply Ruppeiner state space geometry analysis.   
\subsection*{Ehrenfest Scheme of phase transition for the system}
\noindent We check the validity of Ehrenfest equations at the critical points for this black hole system too. Left hand side of the first Ehrenfest equation in eq.(\ref{w}) at critical points $S_i$ can be calculated using eq.(s)(\ref{z},~\ref{df2},~\ref{df3}) as
\begin{eqnarray}
-\left[\left(\dfrac{\partial \Phi}{\partial T}\right)_S\right]_{S=S_i}=\dfrac{S_i-4\alpha\pi}{Q_i}\left[1+\left(1+\dfrac{Q_i^2\pi^2}{b^2(S_i-4\alpha\pi)^2}\right)^{1/2}{}_2F_1\left[\dfrac{1}{4},\dfrac{1}{2},\dfrac{5}{4},-\dfrac{Q_i^2\pi^2}{b^2(S_i-4\alpha\pi)^2}\right]\right]
\end{eqnarray} 
and right hand side is calculated using eq.(s)(\ref{a4},~\ref{df1},~\ref{df2}) as
\begin{eqnarray}
\dfrac{\Delta C_{\Phi}}{T_i Q_i\Delta\beta}=
=\dfrac{S_i-4\alpha\pi}{Q_i}\left[1+\left(1+\dfrac{Q_i^2\pi^2}{b^2(S_i-4\alpha\pi)^2}\right)^{1/2}{}_2F_1\left[\dfrac{1}{4},\dfrac{1}{2},\dfrac{5}{4},-\dfrac{Q_i^2\pi^2}{b^2(S_i-4\alpha\pi)^2}\right]\right]~
\label{mma}
\end{eqnarray}
which agree with each other.\

\noindent Using eq.(\ref{a7}), the left hand side of second Ehrenfest equation at critical points becomes\

\begin{eqnarray}
-\left[\left(\dfrac{\partial \Phi}{\partial T}\right)_Q\right]_{S=S_{i}}=\dfrac{S_i-4\alpha\pi}{Q_i}\left[1+\left(1+\dfrac{Q_i^2\pi^2}{b^2(S_i-4\alpha\pi)^2}\right)^{1/2}{}_2F_1\left[\dfrac{1}{4},\dfrac{1}{2},\dfrac{5}{4},-\dfrac{Q_i^2\pi^2}{b^2(S_i-4\alpha\pi)^2}\right]\right]~.
\label{e2r}
\end{eqnarray}
The right hand side of second Ehrenfest equation is calculated using eq.(s)(\ref{a13}, \ref{df1}, \ref{df4}) as
\begin{eqnarray}
\dfrac{\Delta \beta}{\Delta \kappa}=\dfrac{S_i-4\alpha\pi}{Q_i}\left[1+\left(1+\dfrac{Q_i^2\pi^2}{b^2(S_i-4\alpha\pi)^2}\right)^{1/2}{}_2F_1\left[\dfrac{1}{4},\dfrac{1}{2},\dfrac{5}{4},-\dfrac{Q_i^2\pi^2}{b^2(S_i-4\alpha\pi)^2}\right]\right]~.
\label{e2l}
\end{eqnarray}
This proves the validity of second Ehrenfest equation too. This validates the second order nature of phase transtion points $S_i$. 
Prigogine-Defay ratio for the black hole can also be calculated using eq(s)(\ref{mma}, \ref{e2l}) as
\begin{eqnarray}
\Pi=\dfrac{\Delta C_{\Phi}\Delta\kappa}{T_i Q_i(\Delta\beta)^2}=1~.
\end{eqnarray}
 We further study Ruppeiner's state space geometry analysis for the black hole as studied in section-II.
\subsection*{State space geometry analysis}
\noindent We calculate the Ruppeiner curvature for this black hole spacetime as done in section-II. Mass relation at unit AdS radius from eq.(\ref{bo5}) is
\begin{eqnarray}
M=\dfrac{1}{2}\sqrt{\dfrac{S-4\alpha\pi}{\pi}}+\dfrac{1}{2}\left(\dfrac{S-4\alpha\pi}{\pi}\right)^{3/2}+\dfrac{b^2}{3}\left(\dfrac{S-4\alpha\pi}{\pi}\right)^{3/2}\left(1-\sqrt{1+\dfrac{Q^2\pi^2}{b^2(S-4\alpha\pi)^2}}\right)\nonumber\\
 +\dfrac{2Q^2}{3}\sqrt{\dfrac{\pi}{S-4\alpha\pi}}{}_2F_1\left[\dfrac{1}{4},\dfrac{1}{2},\dfrac{5}{4},-\dfrac{Q^2\pi^2}{b^2(S-4\alpha\pi)^2}\right]~.
\end{eqnarray} 
Ruppeiner metric coefficients are calculated from above equation and these take the form
\begin{eqnarray}
g_{QQ}&=&\dfrac{1}{2T}\sqrt{\dfrac{\pi}{S-4\alpha\pi}}\left[\left(1+\dfrac{Q^2\pi^2}{b^2(S-4\alpha\pi)^2}\right)^{-1/2}+{}_2F_1\left[\dfrac{1}{4},\dfrac{1}{2},\dfrac{5}{4},-\dfrac{Q^2\pi^2}{b^2(S-4\alpha\pi)^2}\right]\right]~\nonumber\\
g_{SQ}&=&-\dfrac{1}{2T}\sqrt{\dfrac{\pi}{S-4\alpha\pi}}\dfrac{Q}{(S-4\alpha\pi)}\left(1+\dfrac{Q^2\pi^2}{b^2(S-4\alpha\pi)^2}\right)^{-1/2}~\nonumber\\
g_{SS}&=&\dfrac{1}{8T}\sqrt{\dfrac{\pi}{S-4\alpha\pi}}\left[-\dfrac{1}{\pi(S-4\alpha\pi)}+\dfrac{3}{\pi^2}+\dfrac{4b^2}{\pi^2}\left(1-\sqrt{1+\dfrac{Q^2\pi^2}{b^2(S-4\alpha\pi)^2}}\right)+\dfrac{4Q^2}{(S-4\alpha\pi)^2}\left(1+\dfrac{Q^2\pi^2}{b^2(S-4\alpha\pi)^2}\right)^{-1/2}\right]~.
\label{a22}
\end{eqnarray}
We can now calculate the Ruppeiner curvature and plot it with entropy for same parameters which shows same discontinuities. 
 \begin{figure}[!ht]
		\subfloat[\label{f1}]{{\includegraphics[scale=.7]{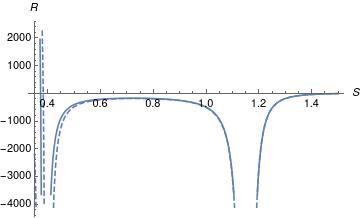}}}
		\qquad
		~~~~~~~~~~
		\caption{Ruppeiner curvature vs. Entropy}{Solid line (Q=0.2, $\alpha$=0.02, b=10)\\
		Dashed line (Q=0.2, $\alpha$=0.02, b$\rightarrow\infty$)}
\end{figure}
This conforms the second order nature of the phase transition. Effects of non-linear electromagnetism are again certified showing same nature as in heat capacity plot.

\section{Conclusion}
 
\noindent In this work, we analysed the black holes with Gauss-Bonnet correction term to the curvature in AdS spacetime. We studied these black holes with charge due to Maxwell and Born-Infeld modified non-linear electromagnetism. Gauss-Bonnet lagrangian does not change the dynamics in $3+1-$ spacetime dimensions, however, black hole entropy gets modified. We calculated the thermodynamic properties associated with the spacetime and calculated heat capacity which showed infinite discontinuities when plotted with entropy at constant other parameters. We further analysed these singularities. Smoothness of $T-S$ plots ruled out any first order phase transition in both the cases. We employed Ehrenfest scheme and Ruppeiner geometry analysis to identify the order of phase transition. Considered systems followed two Ehrenfest equations certifying second order nature of phase transition. We reconfirmed this using Ruppeiner curvature calculations. We calculated and plotted Ruppeiner curvature with entropy for same fixed parameters. Singularities are observed exactly at those points where heat capacity diverged. Both these methods reconfirms second order phase transition. We conclude that the effects of Gauss-Bonnet parameter on the nature of thermodynamic quantities as well as on phase transition points are merely quantitative as compared to black holes in Einstein gravity. The Born-Infeld parameter also shifts the points of phase transitions, however, the effect dominates only in low entropy values.

\section*{Appendix: Simplified Ehrenfest equations}
\noindent In this Appendix, we shall recast the Ehrenfest equations in a simplified form whose validity we have checked in the main text to establish the second order nature of the black hole phase transitions.\

\noindent The left hand side of the first Ehrenfest equation (\ref{w}) at the critical point can be written as
\begin{eqnarray}
-\left[\left(\dfrac{\partial \Phi}{\partial T}\right)_S\right]_{S=S_i}=-\dfrac{\left[\left(\dfrac{\partial \Phi}{\partial Q}\right)_{S}\right]_{S=S_i}}{\left[\left(\dfrac{\partial T}{\partial Q}\right)_{S}\right]_{S=S_{i}}}~.
\label{z}
\end{eqnarray}
\noindent Also, using eq.(\ref{y}) and the definition of heat capacity $C_{\Phi}= T(\frac{\partial S}{\partial T})_{\Phi}$, the right hand side of eq.(\ref{w}) becomes 
\begin{eqnarray} 
Q_i\beta=\left[\left(\dfrac{\partial Q}{\partial T}\right)_{\Phi}\right]_{S=S_i}=\left[\left(\dfrac{\partial Q}{\partial S}\right)_{\Phi}\right]_{S=S_i}\left(\dfrac{C_{\Phi}}{T_i}\right)
\label{a2}
\end{eqnarray}
which then yields
\begin{eqnarray}
\dfrac{\Delta C_{\Phi}}{T_i Q_i\Delta\beta}=\left[\left(\dfrac{\partial S}{\partial Q}\right)_{\Phi}\right]_{S=S_i}~.
\label{a3}
\end{eqnarray}
\noindent Using the identity (\ref{n}), the above equation can be written in the form
\begin{eqnarray}
\dfrac{\Delta C_{\Phi}}{T_i Q_i\Delta\beta}=-\dfrac{\left[\left(\dfrac{\partial \Phi}{\partial Q}\right)_S\right]_{S=S_i}}{\left[\left(\dfrac{\partial \Phi}{\partial S}\right)_Q\right]_{S=S_i}}~.
\label{a4}
\end{eqnarray}
The second Ehrenfest equation can also be recast in a form such that both sides of the equation can be computed easily to check its validity. For that we use the thermodynamic relation $T\equiv T(S,\Phi)$, which gives 
\begin{eqnarray}
\left(\dfrac{\partial T}{\partial \Phi}\right)_Q=\left(\dfrac{\partial T}{\partial S}\right)_{\Phi}\left(\dfrac{\partial S}{\partial \Phi}\right)_Q+\left(\dfrac{\partial T}{\partial \Phi}\right)_S~.
\label{a6}
\end{eqnarray}
Since heat capacity diverges at the critical points, hence $\left[\left(\dfrac{\partial T}{\partial S}\right)_{\Phi}\right]_{S=S_i}=0$, and $\left(\dfrac{\partial S}{\partial \Phi}\right)_Q$ is finite at the critical point, therefore the above equation becomes
\begin{eqnarray}
\left[\left(\dfrac{\partial T}{\partial \Phi}\right)_Q\right]_{S=S_i}=\left[\left(\dfrac{\partial T}{\partial \Phi}\right)_S\right]_{S=S_i}
\label{a7}
\end{eqnarray}
\noindent Now from eq.(\ref{y}), at the critical points
\begin{eqnarray}
\kappa Q_i=\left[\left(\dfrac{\partial Q}{\partial \Phi}\right)_T\right]_{S=S_i}~.
\label{a10}
\end{eqnarray}
Using the thermodynamic identity $\left(\dfrac{\partial Q}{\partial \phi}\right)_T\left(\dfrac{\partial \Phi}{\partial T}\right)_Q\left(\dfrac{\partial T}{\partial Q}\right)_{\Phi}=-1$ and the definition of $\beta$, we find
\begin{eqnarray}
\kappa Q_i=\left[\left(\dfrac{\partial T}{\partial \Phi}\right)_Q\right]_{S=S_i}Q_i\beta~.
\label{a11}
\end{eqnarray}
\noindent Therefore, the right hand side of the second Ehrenfest equation (\ref{x}) reduces to 
\begin{eqnarray}
\dfrac{\Delta \beta}{\Delta \kappa}=-\left[\left(\dfrac{\partial \Phi}{\partial T}\right)_Q\right]_{S=S_i}~.
\label{a12}
\end{eqnarray}
This can further be written as
\begin{eqnarray}
-\left[\left(\dfrac{\partial \Phi}{\partial T}\right)_Q\right]_{S=S_i}=-\dfrac{\left[\left(\dfrac{\partial \Phi}{\partial S}\right)_Q\right]_{S=S_i}}{\left[\left(\dfrac{\partial T}{\partial S}\right)_Q\right]_{S=S_i}}~.
\label{a13}
\end{eqnarray}

 \def\bibsection{\section*{\refname}} 

 \end{document}